\documentclass[aps,twocolumn,showpacs,prb,floatfix]{revtex4-1}
\usepackage{bm}
\usepackage{graphicx}
\usepackage{amsmath,amssymb}
\usepackage{float}
\usepackage{wrapfig}
\usepackage{layout}
\usepackage{color}

\newcommand{\cc}{\mathrm{c}}
\newcommand{\n}{\mathrm{n}}
\newcommand{\BdG}{\mathrm{BdG}}
\newcommand{\imp}{\mathrm{imp}}
\newcommand{\ii}{\mathrm{i}}
\newcommand{\F}{\mathrm{F}} 
\newcommand{\trans}{{}^{\mathrm{t}}}

\newcommand{\Tr}{\mathrm{Tr}}

\begin{document}
\preprint{aps/}
\title{Impurity Effects on Caroli--de Gennes--Matricon Mode in Vortex Core in Superconductors}

\author{Yusuke Masaki}
\email{masaki@vortex.c.u-tokyo.ac.jp}
\affiliation{Department of Physics, The University of Tokyo, Tokyo 113-0033, Japan}
\author{Yusuke Kato}
\affiliation{Department of Basic Science, The University of Tokyo, Tokyo 153-8902, Japan}

\date{\today}
\begin{abstract}
We have developed a scheme of Gor'kov Green's functions to treat the impurity effects on the Caroli--de Gennes--Matricon (CdGM) mode in superconductors (SCs) by improving the Kopnin--Kravtsov scheme with respect to the coherence factors and applicability to various SCs. 
We can study the impurity effects while keeping the discreteness of the energy spectrum, 
in contrast to the quasiclassical theory.
We can thus apply this scheme to SCs with a small quasiclassical parameter $k_{\F}\xi_{0}$ (which is the product of the Fermi wavenumber $k_{\F}$ and the coherence length $\xi_{0}$ in a pure SC at zero temperature) and/or in the superclean regime $\Delta_{\mathrm{mini}} \tau \gg1$ ($\Delta_{\mathrm{mini}}$ and $\tau$ denote, respectively, the level spacing of the CdGM mode called the minigap and the relaxation time for the CdGM mode and we take $\hbar =1$).
We investigate the impurity effects as a white noise for a vortex in an s-wave SC and two types of vortex in a chiral p-wave SC for various values of the quasiclassical parameter and impurity strength 
(from the moderately clean regime to the superclean regime) and confirm the validity of this scheme.
\end{abstract}
\maketitle

\section{Introduction}
In a vortex core in a superconductor (SC), there are electronic bound states called the Caroli--de Gennes--Matricon (CdGM) mode, which has been predicted by Caroli {\it et\ al.} \cite{Caroli1964} and observed experimentally \cite{PhysRevLett.62.214} and  numerically \cite{PhysRevLett.65.1820}. 
These excitations are described as the solutions to the Bogoliubov--de Gennes (BdG) equation. 
The CdGM mode makes an important contribution to phenomena in low-energy physics in vortex states, such as flux flow states \cite{PhysRevB.59.14663, PhysRevB.51.15291}. 
The flux flow conductivity in a dirty SC is often explained by Bardeen-Stephen theory  \cite{PhysRev.140.A1197}, in which the vortex core region is treated as a normal state. In a clean SC, however, the impurity scattering of the CdGM mode is the main mechanism of dissipation in the flux flow states. Moreover, 
particle-hole asymmetry in the CdGM mode causes a flux flow Hall effect. The impurity effects on the CdGM mode play a key role in transport properties in mixed states.

Recently this vortex core mode has attracted much attention owing to the existence of Majorana quasiparticles in zero-energy states in topological SCs \cite{RevModPhys.82.3045, RevModPhys.83.1057}.
The Majorana quasiparticle is its own antiparticle and is expected to exist at the edge or at topological defects such as a vortex core of a topological SC.
An idea for the realization of topological quantum computation (TQC)  \cite{PhysRevLett.86.268} has been proposed, which makes use of non-Abelian statistics in the Majorana zero-energy states in vortex cores.
However, it requires adiabatic braiding operations of Majorana zero-energy states. 
This adiabatic time scale should be less than  the inverse of the minigap $\Delta_{\mathrm{mini}}$, which is the level spacing of the CdGM mode in the vicinity of zero energy (in this paper we take $\hbar = 1$). 
Typically the minigap is on the order of $ \Delta_{\mathrm{mini}}\sim\Delta_{0}/(k_{\F}\xi_{0})$ where $\Delta_{0}$, $k_{\F}$,  and $\xi_{0}$ are, respectively, the modulus of the pair potential in the bulk far from the vortex core, the Fermi wavenumber,  and the coherence length at zero temperature. In $\text{Sr}_{2}\text{RuO}_{4}$, which is expected to be a chiral p-wave SC, $\Delta_{\mathrm{mini}}$ is estimated to be on the order of $1 \mathrm{mK}$  \cite{MaenoKittakaNomuraYonezawaIshida2012}. 
The magnitude of the minigap is relevant for the realization of physics associated with Majorana states.
Since real devices include impurities, it is an important issue  how randomness affects the minigap and the possibility of TQC. 

There are three regimes for the inverse of the relaxation time $\tau_{\n}^{-1}$: 
(i) the dirty regime $\tau_{\n}^{-1}>\Delta_{\mathrm{bulk}}$, 
(ii) the moderately clean regime $\Delta_{\mathrm{mini}}< \tau_{\n}^{-1} \ll \Delta_{\mathrm{bulk}}$, 
(iii) the superclean regime $\tau_{\n}^{-1}\ll\Delta_{\mathrm{mini}}$. 
It is reported that impurity effects depend on the number of impurities $N_{\mathrm{imp}}$ in a vortex core as well as the relaxation time $\tau_{\n}$, particularly in the superclean regime. 
In other words, $\tau_{\n}$ can be expressed as $\Delta_{\mathrm{mini}}\tau_{\n} \propto (N_{\mathrm{imp}}\theta^{2})^{-1}$, 
where $\theta$ is the Born parameter and characterizes the type of scatterers, 
and the magnitudes of both $N_{\mathrm{imp}}$ and $\theta$ are crucial for determining impurity effects, 
especially for large $\tau_{\n}$. 

In the presence of a few impurities with the Born parameter being not small, $1/\sqrt{k_{\F}\xi}\ll\theta\lesssim 1$, 
the $2\Delta_{\mathrm{mini}}$-periodic structure of the energy spectrum appears 
rather than the $\Delta_{\mathrm{mini}}$-periodic one~\cite{PhysRevB.57.5457,PhysRevB.59.12021,PhysRevB.60.14597}.
These kinds of impurity always give rise to level mixing, and the Landau-Zener transition causes the dominant absorption in flux flow states in the (moderately clean and) superclean regimes.
For an unconventional superconductor such as a chiral p-wave SC or d-wave SC, 
these impurities lead to the density of states (DOS) 
around the Fermi energy, corresponding to the midgap bound states~\cite{RevModPhys.78.373}.

The adiabatic dynamics in TQC requires, however, exceptionally small absorption, making 
it also important to consider the nondissipative regime for vanishingly small $\theta$. 
We assume there are weak but many scatterers in a vortex core. 
This kind of impurity is called white noise disorder.
In the dirty regime and moderately clean regime in which the Born parameter is very small, 
analysis with random matrix theory (RMT) has clarified 
that the level statistics obey class C of the Altland-Zirnbauer classification~\cite{PhysRevB.55.1142,SkvortsovFeigelmanKravtsov1998,PhysRevB.62.15190}. 
Impurity scattering rates have been obtained by analysis using the quasiclassical theory~\cite{PhysRevB.51.15291,Kato2000}. In these studies, spectra form peaks with widths. 
The analysis by the quasiclassical theory has also clarified 
the difference in the types of SC, such as s-wave and chiral p-wave SCs. 
The difference stems from coherence factors, 
and the cancellation of coherence factors corresponding to  Anderson's theorem 
in the vortex core occurs in one of the vortices in a chiral p-wave SC~\cite{JPSJ.71.1721}.

For the adiabaticity in TQC, we consider the superclean regime with white noise disorder. 
In this regime, we need to treat discrete levels, hence the quasiclassical theory is not applicable.
Kopnin and Kravtsov have developed a scheme of Gor'kov Green's functions with an impurity self-energy to investigate the flux flow Hall effect 
in a three-dimensional type-II s-wave SC~\cite{kopnin1976conductivity}, where they have treated  the vortex core states while keeping the discreteness of their levels. 
The form of the coherence factors and thus their way of application to other SCs are, however, not clear in their scheme.
In this paper, to overcome these difficulties, we derive a suitable scheme for Gor'kov Green's functions with an impurity self-energy in two dimensions 
and apply it to a chiral p-wave SC with two types of vortex as well as an s-wave SC with a single vortex. 
This scheme is easily extended to three-dimensional cases if we do not consider numerical costs. 
We confirm the validity of this scheme 
by comparing the results for impurity effects in a system with $k_{\F}\xi_{0} = 100$ with those of the quasiclassical theory.
We also investigate the impurity effects in the large-$\Delta_{\mathrm{mini}}\tau$ regime ($\tau$ denotes the relaxation time for the CdGM mode), where the system is not in the quasiclassical regime, 
and find that the impurities affect spectra in the low-energy region in a way similar to those in the quasiclassical limit, except for the zero-energy state.

In the following, we use the unit $k_{\mathrm{B}}=1$ in addition to $\hbar = 1$ for convenience.

\section{Method} \label{method}
\subsection{General approach} \label{general-approach}
We consider the Gor'kov equation for superconductors with a single vortex and rotational symmetry. 
The system has a disc geometry of radius $R$ from the vortex core at the origin.
The Gor'kov equation in real space is transformed into the Dyson equation in real space:
\begin{align}
\check{G}(\bm{r},\bm{r}';\omega_{n}) 
&= \check{G}^{(0)}(\bm{r},\bm{r}';\omega_{n}) \nonumber \\
+\int d\bm{r}_1
&\check{G}^{(0)}(\bm{r},\bm{r}_1;\omega_{n})
\check{\Sigma}_{\mathrm{imp}}(\bm{r}_1; \omega_n)
\check{G}(\bm{r}_1,\bm{r}';\omega_{n}), \label{gorkov1}
\end{align}
where $\check{\ }$ denotes a $4\times 4$ matrix, $\check{G} (\check{G}^{(0)}) $ Green's function with (without) impurities, and $\check{\Sigma}_{\imp}$ an impurity self-energy.
We define Green's functions by
\begin{align}
\check{G}(x_{1},x_{2}) &= \begin{pmatrix}
\hat{G} (x_1,x_2) & \hat{F} (x_1,x_2) \\
-\hat{F}^\dagger (x_1,x_2) & \hat{\bar{G}} (x_1,x_2) 
\end{pmatrix}.
\label {matrixG}
\end{align}
Each element in Eq.~\eqref{matrixG} represents a 2$\times$2 matrix, the matrix elements of which are given by
\begin{align}
(\hat{G}(x_1,x_2))_{\alpha\beta}&=\langle T\psi_\alpha(x_1)\psi_\beta^\dagger(x_2)\rangle, \label{G1}\\
(\hat{\bar{G}}(x_1,x_2))_{\alpha\beta}&=-\langle T\psi_\alpha^\dagger(x_1)\psi_\beta(x_2)\rangle, \label{G2}\\
(\hat{F}(x_1,x_2))_{\alpha\beta}&=\langle T\psi_\alpha(x_1)\psi_\beta(x_2)\rangle, \label{F}\\
(\hat{F}^\dagger(x_1,x_2))_{\alpha\beta}&=\langle T\psi_\alpha^\dagger(x_1)\psi_\beta^\dagger(x_2)\rangle.\label{Fd}
\end{align}
The symbols $\alpha$ and $\beta$ denote the electron spins $\uparrow$ or $\downarrow$. We introduce the Fourier transformation for relative time $\bar{\tau}\equiv\tau_{1}-\tau_{2}$:
\begin{align}
\check{G}(x_1,x_2)\equiv \check{G}(\bm{r}_1,\bm{r}_2;\bar{\tau})=T \sum_n\check{G}(\bm{r}_1,\bm{r}_2;\omega_{n})e^{-\ii\omega_n\bar{\tau}},
\end{align}
where $\omega_{n}$ denotes the fermionic Matsubara frequency.
Within the self-consistent Born approximation (SCBA), $\check{\Sigma}_{\imp}$ is described as
\begin{align}
\check{\Sigma}_{\imp}(\bm{r};\omega_{n}) = \dfrac{\Gamma_{\n}}{\pi N_{\n}} \check{G}(\bm{r},\bm{r};\omega_{n})\equiv \gamma \check{G}(\bm{r},\bm{r};\omega_{n}).
\end{align}
$N_{\n}$ denotes the density of states per spin at the Fermi energy $\epsilon_{\F}$ in the normal state. 
$\Gamma_{\n}$ is the impurity scattering rate in the normal state and related to the relaxation time $\tau_{\n}$ as $\Gamma_{\n}=1/\tau_{\n}$.
$\check{G}^{(0)}$ can be obtained from the BdG equation without impurities:
\begin{align}
\check{H}_{\BdG}^{(0)} (\bm{r})\bm{u}_{K}(\bm{r})=E_{K} \bm{u}_{K}(\bm{r}). \label{BdG}
\end{align}
$\check{H}_{\BdG}^{(0)} (\bm{r})$ is the BdG Hamiltonian for the pure SC.
The basis takes the form $\bm{u}_{K}(\bm{r})=\trans(u_{K\uparrow}(\bm{r}), u_{K\downarrow}(\bm{r}), v_{K\uparrow}(\bm{r}), v_{K\downarrow}(\bm{r}))$ with index $K$.
The expression for $\check{G}^{(0)}$ is given by
\begin{align}
\check{G}^{(0)}(\bm{r},\bm{r}';\omega_{n}) = \sum_{K} \dfrac{\check{\tau}_{z}\bm{u}_{K}(\bm{r})(\bm{u}_{K}(\bm{r}'))^{\dagger}}{E_{K}-\ii\omega_{n}}. \label{g0}
\end{align}
Since the system without impurities has rotational symmetry, the modified angular momentum $\check{J}_{z}$,
\begin{align}
\check{J}_{z} = L_{z}\check{1} +l_{1}\check{\sigma}_{z}\check{\tau}_{z}+l_{2}\check{\tau}_{z},
\end{align}
commutes with the Hamiltonian $\check{H}_{\BdG}^{(0)}$, and its eigenvalue $l$ is a good quantum number. We define a unitary operator 
$\check{U}_{l}(\theta)= \mathrm{diag}(e^{i(l-l_{1}-l_{2})\theta}, e^{i(l+l_{1}-l_{2})\theta}, e^{i(l+l_{1}+l_{2})\theta}, e^{i(l-l_{1}+l_{2})\theta})/\sqrt{2\pi}$, 
and describe the eigenvector $\bm{u}_{K}(\bm{r})$ as
\begin{align}
\bm{u}_{K}(\bm{r}) = \check{U}_{l}(\theta) \bm{\mathcal{R}}_{l,\nu}(r).
\end{align}
Here we introduce $(r,\theta)$ as two-dimensional polar coordinates. 
The basis is described as $(\bm{e}_{r},\bm{e}_{\theta}) = R(\theta) (\bm{e}_{x},\bm{e}_{y})$, where $R(\theta)$ is a $2 \times 2$ matrix that rotates the vector around the $z$-axis by angle $\theta$. 
The index $K$ is rewritten as the set of the angular momentum $l$ and the quantum number of the radial part $\nu$. 
To eliminate the $\theta$ dependence from the BdG equation, we multiply both sides of Eq.~\eqref{BdG} by $\check{U}_{l'}^{\dagger}(\theta)$ from the left  
and integrate them with respect to $\theta$ over $0$ to $2\pi$.
The BdG equation for this radial part $\bm{\mathcal{R}}_{l,\nu}(r)$ is obtained as
\begin{align}
\check{\mathcal{H}}^{(0)}_{l}(r) \bm{\mathcal{R}}_{l,\nu}(r) = E_{l,\nu}\bm{\mathcal{R}}_{l,\nu}(r),
\end{align} 
where $\int_{0}^{2\pi}d\theta \check{U}_{l'}^{\dagger}(\theta) \check{H}_{\BdG}^{(0)}(\bm{r}) \check{U}_{l}(\theta)\equiv  \delta_{l,l'}\check{\mathcal{H}}^{(0)}_{l}(r)$. Green's function can be written as 
\begin{align}
\check{G}^{(0)}(\bm{r},\bm{r}';\omega_{n}) 
&=\sum_{l} \check{U}_{l}(\theta)\check{G}_{l}^{(0)}(r,r';\omega_{n})\check{U}_{l}^{\dagger}(\theta'), \\
\check{G}_{l}^{(0)}(r,r';\omega_{n})&= \sum_{\nu} \dfrac{\check{\tau}_{z}\bm{\mathcal{R}}_{l,\nu}(r)(\bm{\mathcal{R}}_{l,\nu}(r'))^{\dagger}}{E_{l,\nu}-\ii\omega_{n}}.
\end{align}
Although impurities break the rotational symmetry, scattering processes do not change the angular momentum 
after averaging over the configurations of impurities. 
This means that Green's function $\check{G}$ also takes the following form:
\begin{align}
\check{G}(\bm{r},\bm{r}';\omega_{n}) 
&=\sum_{l} \check{U}_{l}(\theta)\check{G}_{l}(r,r';\omega_{n})\check{U}_{l}^{\dagger}(\theta'),
\end{align}
and we can obtain the radial part of the Dyson equation as 
\begin{align}
\check{G}_{l}(r,r';\omega_{n})&= \check{G}^{(0)}_{l}(r,r';\omega_{n}) \nonumber \\
+\int r_{1}dr_{1}&\check{G}^{(0)}_{l}(r,r_{1};\omega_{n})
\check{\Sigma}_{\imp}(r_{1};\omega_{n})\check{G}_{l}(r_{1},r';\omega_{n}), \label{Dyson1}\\
\check{\Sigma}_{\imp}(r_{1};\omega_{n})&=\dfrac{\gamma}{2\pi}\sum_{l}\check{G}_{l}(r_{1},r_{1};\omega_{n}). \label{Self1}
\end{align}
We define the scalar quantities
\begin{align}
\sigma_{l,\nu,\nu'}(\ii \omega_n) = \int rdr\bm{\mathcal{R}}_{l,\nu}^{\dagger}(r)\dfrac{\check{\Sigma}_{\imp}(r;\omega_n)
\check{\tau}_{z}\bm{\mathcal{R}}_{l,\nu'}(r)}{E_{l,\nu'}-\ii\omega_{n}}, \label{generalscalar}
\end{align}
and they constitute a matrix $\bar{\sigma}_{l}(\ii \omega_{n})$: $(\bar{\sigma}_{l}(\ii \omega_{n}))_{\nu,\nu'}=\sigma_{l,\nu,\nu'}(\ii \omega_n)$.
With this modified self-energy $\bar{\sigma}$, the Dyson equation can be solved formally, and Green's function is obtained as
\begin{align}
\check{G}_{l}(r,r';\omega_n) 
&=\sum_{\nu \nu'}\frac{\check{\tau}_z{\bm{\mathcal{R}}}_{l,\nu}(r)}{E_{l\nu}-\ii\omega_{n}}\left[\sum_{k}\left(\bar{\sigma}_{l}(\ii \omega_{n})\right)^{k}\right]_{\nu\nu'}
\bm{\mathcal{R}}^{\dagger}_{l,\nu'}(r') \nonumber \\
&=\sum_{\nu\nu'}\frac{\check{\tau}_z\bm{\mathcal{R}}_{l,\nu}(r)}{E_{l\nu}-\ii\omega_{n}}\left[(\bar{1}-\bar{\sigma}_l(\ii \omega_n))^{-1}\right]_{\nu\nu'}
\bm{\mathcal{R}}^{\dagger}_{l,\nu'}(r'). \label{greenself}
\end{align}
Substituting Eqs. \eqref{Self1} and \eqref{greenself} into Eq.~\eqref{generalscalar}, we obtain an equation for $\bar{\sigma}$:
\begin{align}
\sigma_{l,\nu_1,\nu_2}(\ii \omega_n)=
\dfrac{\gamma}{2\pi} \sum_{l',\nu_1',\nu_2'}\dfrac{M^{ll'}_{\nu_1\nu_2\nu_1'\nu_2'}\left[(\bar{1}-\bar{\sigma}_{l'}(\ii \omega_n))^{-1}\right]_{\nu_1'\nu_2'}}
{(\ii\omega_n-E_{l'\nu_1'})(\ii\omega_n-E_{l\nu_2})}, \label{genselfconsistent}
\end{align}
where 
\begin{align}
M^{ll'}_{\nu_1\nu_2\nu_1'\nu_2'}=\int r dr [\bm{\mathcal{R}}_{l,\nu_1}^\dagger(r)\check{\tau}_z\bm{\mathcal{R}}_{l',\nu_1'}(r)]
[\bm{\mathcal{R}}_{l',\nu_2'}^\dagger(r)\check{\tau}_z\bm{\mathcal{R}}_{l,\nu_2}(r)], \label{genmatele}
\end{align}
and these matrix elements [Eq.~\eqref{genmatele}] represent coherence factors.
The density of states can be evaluated from the Green's functions obtained above as follows: 
\begin{align}
N(\omega) 
&\equiv \int d\bm{r} \mathrm{Im}\left[\dfrac{1}{\pi}\left.\Tr \check{\tau}_{z}\check{G}(\bm{r},\bm{r};\omega_{n})\right|_{\ii\omega_{n}\to\omega+i\delta}\right] \nonumber \\
&=\sum_{l,\nu}\mathrm{Im}\left[ \dfrac{-\left[(\bar{1}-\bar{\sigma}_{l}(\omega+i\delta))^{-1}\right]_{\nu\nu}}{\pi(\omega+i\delta-E_{l\nu})}\right] \equiv \sum_{l,\nu}N_{l,\nu}(\omega). \label{defDOS}
\end{align}
\subsection{Single-mode approximation} \label{single-mode-approximation}
We are interested in the impurity effects on the CdGM mode and assume that the main contribution to this mode is from the CdGM mode itself.
We thus restrict $\nu$ to that for the CdGM mode, to which we assign $\nu=0$. We will examine the validity of this restriction in Sect.~\ref{bulk-mode}. We redefine $\sigma$ and $M$ as follows:
\begin{align}
\sigma_{l}(\ii \omega_{n}) &\equiv(E_{l,0}-\ii\omega_{n})\sigma_{l,0,0}(\ii \omega_{n}), \label{defsinsigma} \\
M_{l,l'}&\equiv M^{ll'}_{0,0,0,0}. \label{defsinmat}
\end{align}
Equations \eqref{genselfconsistent} and \eqref{genmatele} are then reduced to
\begin{align}
\sigma_{l}(\ii \omega_n)&=\dfrac{\gamma}{2\pi} \sum_{l'}\dfrac{M_{l,l'}}{E_{l'0}-\sigma_{l'}(\ii \omega_n)-\ii\omega_n}, \label{sinselfconsistent} \\
M_{l,l'}&=\int r dr |\bm{\mathcal{R}}_{l,0}^\dagger(r)\check{\tau}_z\bm{\mathcal{R}}_{l',0}(r)|^{2}, \label{sinmatele}
\end{align}
and Green's function $\check{G}_{l}$ is rewritten as 
\begin{align}
\check{G}_{l}(r,r';\omega_{n}) =   \dfrac{\check{\tau}_{z}\bm{\mathcal{R}}_{l, 0}(r)(\bm{\mathcal{R}}_{l, 0}(r'))^{\dagger}}{E_{l, 0}-\sigma_{l}(\ii \omega_{n})-\ii\omega_{n}}. \label{singreen}
\end{align}
Hereafter, we call this scheme the improved Kopnin--Kravtsov (iKK) scheme. 
The iKK scheme is advantageous to the scheme derived by Kopnin and Kravtsov  \cite{kopnin1976conductivity};
Eq. \eqref{sinselfconsistent} has the coherence factor $M_{l,l'}$, which was not clear in  Eq.~(1) of Ref.~\onlinecite{kopnin1976conductivity}. 
The coherence factor depends on the pairing symmetry, and the iKK scheme can thus reflect differences among various types of SC.
Since Eqs.~\eqref{sinselfconsistent}--\eqref{singreen} are constructed with the eigensystems of the BdG equation formally without reference to their analytic expressions,
we can easily apply this method to other superconducting systems with a single vortex for which an analytical solution to the CdGM mode is not available.

\begin{figure}[t]
\begin{center}
\includegraphics[height = 5cm]{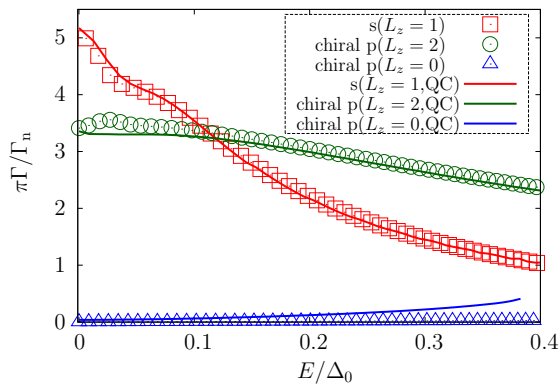}
\caption{(Color online) Impurity scattering rate $\Gamma$ as a function of energy eigenvalue $E$ for $L_z=0,1,2$. QC means the quasiclassical calculation and solid lines are calculated by the quasiclassical theory \cite{JPSJ.71.1721}. Except for the case with $L_{z} = 0$, the results of the iKK scheme are in good agreement with those of the quasiclassical theory. }
\label{widsp100v2v1}
\end{center}
\end{figure}

\begin{figure*}[t]
\begin{center}
\includegraphics[width=17cm]{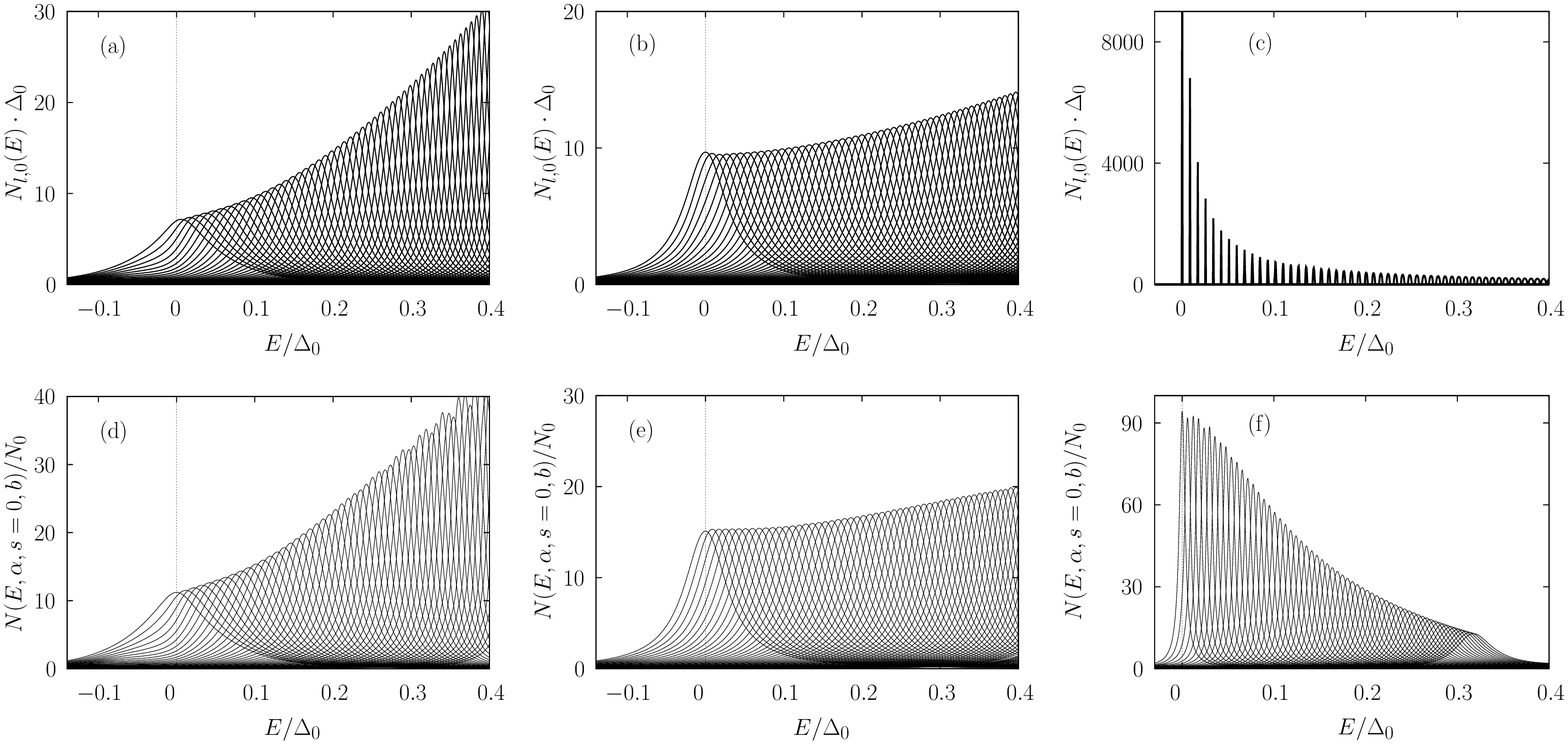}
\end{center}
\caption{DOS for CdGM modes of the s-wave SC ((a) (d), $L_{z}=1$) and the chiral p-wave SCs with the parallel vortex ((b) (e), $L_{z}=2$) and antiparallel vortex ((c) (f), $L_{z}=0$). (a)--(c) are obtained by the iKK scheme and (d)--(f) are obtained by the quasiclassical theory \cite{JPSJ.71.1721}. The quasiclassical parameter of (a)--(c) is taken as $k_{\F}\xi_{0} = 100 $.  The normal scattering rates are taken as $\Gamma_{\mathrm{n}} = 10^{-2}\pi \Delta_{0}$ for $L_{z}=1, 2$ and $\Gamma_{\mathrm{n}} = 10^{-1}\pi \Delta_{0}$ for $L_{z}=0$.
}
\label{dos100v2}
\end{figure*}
\section{Numerical Study}\label{numerical-study}
We consider two cases, where the quasiclassical parameter $k_{\F}\xi_{0}$ is $100$ and $5$. We calculate the DOS within the single-mode approximation introduced in Sect.~\ref{single-mode-approximation}. 
We take the units of energy and length for $\Delta_{0}$ and $\xi_{0}$, which are the modulus of the pair potential in the bulk far from the vortex core and the coherence length defined as
$\xi_{0} = v_{\F}/\Delta_{0}$, respectively. 
In this paper, we consider a single vortex in an s-wave SC and two types of vortex of chiral p-wave SCs. In chiral p-wave SCs, there are two types of vortex, each of which cannot be transformed into the other by any symmetry operation. 
This kind of inequivalence is inherent to SCs in which the time-reversal symmetry is broken even in the absence of a magnetic field; recall that a vortex and antivortex in an s-wave SC are transformed to each other via time-reversal operation. 
Each type of vortex in chiral p-wave SCs is specified by the relative sign of the vorticity and chirality. 

For s-wave and chiral p-wave SCs, the $4 \times 4$ BdG Hamiltonians can be reduced  to the spinless representation. The explicit forms are given by
\begin{align}
\begin{pmatrix}
\dfrac{-\bm{\nabla}^{2}}{2m} -\mu & D(\bm{r}) \\
\bar{D}(\bm{r}) &\dfrac{\bm{\nabla}^{2}}{2m} +\mu 
\end{pmatrix} 
 \begin{pmatrix}
 u_{K}(\bm{r}) \\ v_{K}(\bm{r})
 \end{pmatrix}
=E_{K}
 \begin{pmatrix}
 u_{K}(\bm{r}) \\ v_{K}(\bm{r})
 \end{pmatrix},
 \end{align}
 where \cite{matsumoto1999chiral}
 \begin{align}
 D(\bm{r}) = 
 \begin{cases}
 \Delta(\bm{r})  =\left(\bar{D}(\bm{r})\right)^{*} \hspace{3em} \text{s-wave},\\
\dfrac{1}{\ii k_{\F}}\left\{ \dfrac{1}{2}\left[\partial_{\pm}\Delta(\bm{r})\right]+\Delta(\bm{r})\partial_{\pm}\right\}=-\left(\bar{D}(\bm{r})\right)^{*} \\ 
\partial_{\pm} = \partial_{x}\pm\ii\partial_{y}\hspace{3.5em}\text{chiral p-wave}. 
 \end{cases} \label{pairing}
 \end{align}
 We assume the spatial profile of the pair potential to be
 \begin{align}
 \Delta(\bm{r}) = \Delta_{0}\tanh(r/\xi_{0})e^{\pm \ii \theta}. \label{delta}
 \end{align}
 We take the sign of the chirality in the chiral p-wave SC in Eq.~\eqref{pairing} as positive. For the  chiral p-wave SC, the positive sign in Eq.~\eqref{delta} means that the vorticity is parallel to the chirality
 and the negative sign means that the vorticity antiparallel to the chirality. The former is labeled as $L_{z}=2$ and the latter as $L_{z}=0$, and we call 
 the vortex with $L_{z}=2$ ($L_{z}=0$) the ``parallel (antiparallel) vortex''. 

In order to obtain eigenvalues and eigenfunctions, we use the Fourier-Bessel expansion \cite{PhysRevB.43.7609, matsumoto1999chiral}. We use the radius $R=10\xi_{0}$ 
and two cutoffs, $l_{\cc}=100$, $30$, and $N_{\cc}=200$ for $l$ and $\nu$, respectively. $N_{\cc}$ denotes the number of zero points 
of the Bessel functions we have used.
In the following calculations, we consider $T=0$ and replace $\ii \omega_{n}$ with $\omega+\ii\delta$ in the above scheme represented 
by Eqs.~\eqref{sinselfconsistent}--\eqref{singreen}. 
$\delta$ is  infinitesimal and set to $10^{-5}\Delta_{0}$. 
Impurity effects are evaluated through the width of the spectrum, denoted as $\Gamma$. We define $\Gamma$ as the half width at half maximum, which we also call the impurity scattering rate. The relaxation time $\tau$ for the CdGM mode is associated with $\Gamma$ as $\tau \sim 1/\Gamma$.

\begin{figure*}
\begin{center}
\includegraphics[width=17cm]{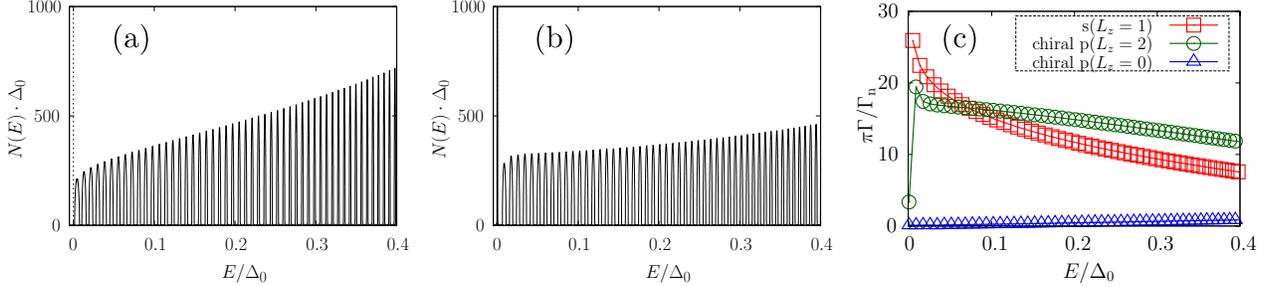}
\end{center}
\caption{(Color online) Impurity effects for the normal scattering rate $\Gamma_{\mathrm{n}} =10^{-4}\pi\Delta_{0}$ and the quasiclassical parameter $k_{\F}\xi_{0} = 100$. (a) and (b) show the DOS for the s-wave SC and the chiral p-wave SC with the parallel vortex, respectively. The widths of the spectra for $L_{z}=0, 1, 2$ are shown in (c). }
\label{100v4}
\end{figure*}
We remark on how to construct the Majorana state localized around the vortex core in numerical calculations for finite-size systems. 
In chiral p-wave SCs with an infinite system size, there are two zero-energy states with $l=0$. 
Their energies become finite but sufficiently small and opposite in sign to each other when the system size is finite and sufficiently large. We use the labels  $(l=0,\nu = +)$ for the positive-energy state and ($l=0,\nu = -$) 
for the negative-energy state. 
Moreover, both their wave functions simultaneously have weight in the vortex core and near the edge.
In order to treat these states within the single-mode approximation, we define the energy and the state labeled by  $\nu = 0$ and $l=0$ as 
$E_{0,0} = (E_{0,+} \pm E_{0,-})/2$ and $\bm{\mathcal{R}}_{0,0}(r) = [\bm{\mathcal{R}}_{0,+}(r)\pm\bm{\mathcal{R}}_{0,-}(r)]/\sqrt{2}$, respectively. 
We choose the sign of $\pm$ so as to obtain the bound state in the vortex core.

\begin{figure*}
\begin{center}
\includegraphics[width=17cm]{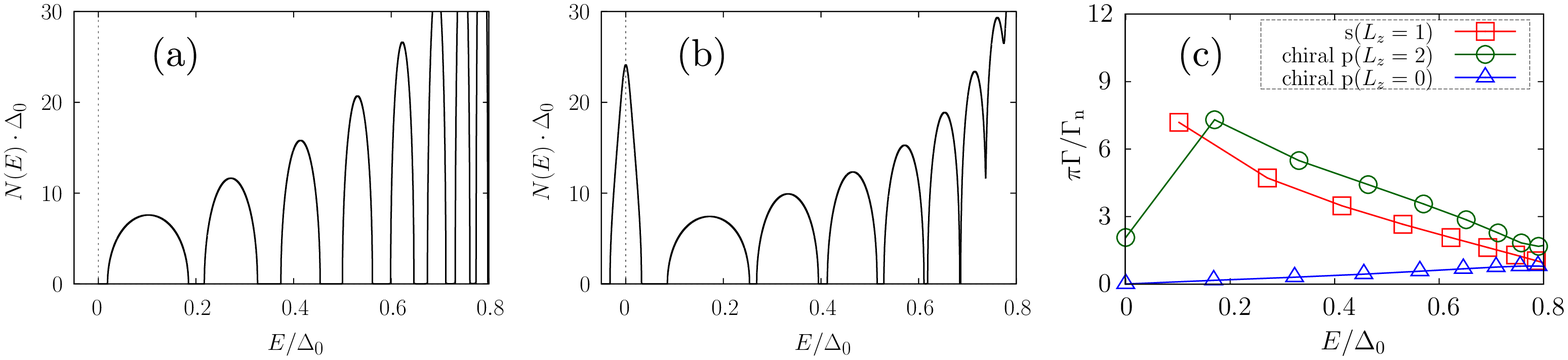}
\end{center}
\caption{(Color online) Impurity effects for the normal scattering rate $\Gamma_{\mathrm{n}} =10^{-2}\pi\Delta_{0}$ and the quasiclassical parameter $k_{\F}\xi_{0} = 5$. (a) and (b) show the DOS for the s-wave SC and the chiral p-wave SC with the parallel vortex, respectively. The widths of the spectra for $L_{z}=0, 1, 2$ are shown in (c).  }
\label{5v2}
\end{figure*}
\subsection{Results: $k_{\mathrm{F}}\xi_{0} = 100$} \label{kfxi100}
For this quasiclassical parameter, we expect that this method will reproduce the result calculated by the quasiclassical theory \cite{JPSJ.71.1721}.
The normal scattering rate $\Gamma_{\n}$ can be changed as a parameter. 
First, we set $\Gamma_{\n}=10^{-2}\pi\Delta_{0}$ for an s-wave SC ($L_{z}=1$) and a chiral p-wave SC with $L_{z}=2$, and $\Gamma_{\n}=10^{-1}\pi\Delta_{0}$ for a chiral p-wave SC with $L_{z}=0$. In these cases, the widths $\Gamma$ are summarized in Fig.~\ref{widsp100v2v1}, which are evaluated from the spectra in Figs.~\ref{dos100v2}(a)--(f).
Figures~\ref{dos100v2}(a)--(c) show the DOS calculated in the iKK scheme for each angular momentum $l$, which takes discrete values. Figures 2(d)--(f) show the angular-resolved local density of states (LDOS)\footnote{
The angular-resolved LDOS means the spectral function that is not integrated with respect to the Fermi momentum on the Fermi surface, and hence
it is different from the conventional LDOS. The detail is explained in Sect. 2 in Ref.~\onlinecite{JPSJ.71.1721}.
 } calculated by the quasiclassical theory for some impact parameters  $b$ ($b/\xi_{0} = 0.01 n, n= 0,1,2,\cdots$), which can take arbitrary values. In Figs.~\ref{dos100v2}(d)--(f), we have introduced another coordinate system spanned by $(\bm{e}_{s},\bm{e}_{b}) = R(\alpha) (\bm{e}_{x},\bm{e}_{y})$, 
where $\alpha$ is the angle between $\bm{k}_{\F}$ and $\bm{e}_{x}$. 
We assume rotational symmetry; thus, the angular-resolved LDOS is independent of the value of $\alpha$ for the fixed set of $(s,b)$.   
We calculate the LDOS at $s=0$, referring to Fig.~3 in Ref.~\onlinecite{JPSJ.71.1721}. 
We remark on the relationship between the DOS and LDOS. In these cases, the angular momentum is approximately described by the impact parameter as $l = -k_{\F} b$, which is not quantized. 
Because of this relationship, the scattering rate and peak energy for the state labeled by $l$ correspond to those for the state labeled by $(\bm{k}_{\F},b)$. Therefore, the DOS and LDOS have similar structures since the scattering rate and peak energy mainly characterize the shape of the spectrum. 

Figure \ref{dos100v2}(a) shows the energy dependence of the DOS $N_{l,0}(E)$ [defined as Eq.~\eqref{defDOS}] for the s-wave SC.
This spectrum is very similar to that obtained by the quasiclassical theory (Fig.~\ref{dos100v2}(d)). 
We find that the spectra also overlap their neighbor states regardless of their discreteness. 
In this parameter region, we can thus consider the system to be in the quasiclassical regime.
In fact, the typical width of the spectrum $\Gamma \sim \Gamma_{\n}= 10^{-2}\pi\Delta_{0}$ is compatible with the level spacing, $\Delta_{0}/(k_{\F}\xi_{0})\sim 10^{-2}\Delta_{0}$; thus, the levels are regarded as continuous.
The DOS and LDOS of the chiral p-wave SC with the parallel vortex ($L_{z}=2$) are shown in Figs.~\ref{dos100v2}(b) and \ref{dos100v2}(e). In this case, each spectrum overlaps its neighbors enough for the system to be in the quasiclassical regime, and 
is in good agreement with that obtained by the quasiclassical theory as for the case of the s-wave SC. The energy dependence of the spectrum broadening is similar to that for the s-wave SC, where the spectrum with lower energy in the CdGM mode broadens more. 

For the chiral p-wave SC with the antiparallel vortex ($L_{z}=0$), the spectra are sharp, as shown in Fig.~\ref{dos100v2}(c), i.e., the spectra are hardly affected by impurities. These small scattering rates are qualitatively the same as those obtained by the quasiclassical theory.
Because of this robustness, there are almost no overlaps among the spectra shown in Fig.~\ref{dos100v2}(c), while in the quasiclassical theory the small overlaps stem from the continuous spectra shown in Fig.~\ref{dos100v2}(f). The widths obtained by this method are slightly different from those obtained by the quasiclassical theory, as can be seen from the blue solid line and blue triangles in Fig.~\ref{widsp100v2v1}. 
For these values of $k_{\F}\xi_{0}$ and $\Gamma_{\n}/\Delta_{0}$, the system with $L_{z}=0$ can no longer be considered to be in the quasiclassical regime.
This implies that we can consider the system is in the quasiclassical regime when there are sufficiently large overlaps among the spectra.
In contrast to the s-wave SC and the chiral p-wave SC with $L_{z}=2$ mentioned above, in terms of the energy dependence of scattering rates, 
spectra around zero energy are more robust against impurities than those at higher energies in CdGM mode. This behavior is qualitatively the same as 
the quasiclassical behavior. 
We thus see that the 
energy dependences of the scattering rates  of CdGM modes are similar for the quasiclassical and quantum ($\Delta_{\mathrm{mini}}\tau \gg1$) limits.

From these figures (Figs.~\ref{dos100v2}(a)--(f)), we thus find that 
the result obtained by the iKK scheme and the quasiclassical theory are in good agreement 
except for the chiral p-wave SC with $L_{z}=0$; 
Figures \ref{dos100v2}(c) and \ref{dos100v2}(f) exhibit a small discrepancy with respect to the overlaps of the spectra. 
The system with $L_{z}=0$ is not in the quasiclassical regime even for $k_{\F}\xi_{0} = 100$, as we mentioned above.

Next, we consider another case, in which the systems are very clean, and hence the widths are small and 
the overlaps among the spectra are also small.  
We consider only the s-wave SC and the chiral p-wave SC with $L_{z}=2$ and set $\Gamma_{\n}=10^{-4}\pi\Delta_{0}$.
Both of these SCs do not have overlaps of the spectra (as shown in Figs.~\ref{100v4}(a) and \ref{100v4}(b)). Actually, the widths $\Gamma$ of the s-wave SC and the chiral p-wave SC with $L_{z} = 2$ are on the order of 
$10^{-3}\Delta_{0}$ (Fig.~\ref{100v4}(c)), while the level spacings are on the order of $10^{-2}\Delta_{0}$.
The tendency of the impurity effects is the same as in the above results and quasiclassical results.
The chiral p-wave SC with $L_{z}=2$ has larger impurity scattering rates at lower energies but the zero-energy state has an exceptionally small scattering rate. 
This is because the matrix element $M_{0,0}$ [Eq.~\eqref{sinmatele}] is zero due to the character of the zero-energy state.
When the spectra are discrete and the contributions from neighbor states are relatively small, this character appears, in contrast to the above state with $L_{z} = 2$.
In this sense, the vortex type of the chiral p-wave SC with $L_{z}=2$ is different from that of the s-wave SC.

\subsection{Results: $k_{\mathrm{F}}\xi_{0} = 5$}\label{kfxi5}
The system with $k_{\F}\xi_{0} = 5$ is not in the quasiclassical regime because of  the large level spacing. 
Let us discuss some differences between the results for the small quasiclassical parameter ($k_{\F} \xi_{0}=5$) and large quasiclassical parameter ($k_{\F}\xi_{0}=100$).  Although the normal scattering rates ($\Gamma_{\n} = 10^{-2}\pi \Delta_{0}$) are the same in these two cases, 
the resultant scattering rates $\Gamma$ in Fig.~\ref{5v2}(c) are larger than those in Fig.~\ref{widsp100v2v1}. 
We infer that a large level spacing leads to a smaller level repulsion, which yields the wider spectra.
The impurity effects thus depend on $k_{\F}\xi_{0}$. This dependence is not accessible in the quasiclassical theory 
because the conventional quasiclassical theory neglects the contributions from $O(1/(k_{\F}\xi_{0}))$.
In Fig.~\ref{5v2}(c), the scattering rate $\Gamma$ is small only for the zero-energy state in the chiral p-wave SC with $L_{z}=2$, as well as that in Fig.~\ref{100v4}(c).
Overall behaviors are qualitatively the same as those in Fig.~\ref{100v4}(c).
We note that since $k_{\F}v_{\F}\tau = k_{\F}\xi_{0}\Delta_{0}/\Gamma_{\n}=500/\pi\gg 1$, weak-localization effects are small, regardless of the small quasiclassical parameter. 

\begin{figure}[t]
\begin{center}
\includegraphics[height=7cm,angle =270]{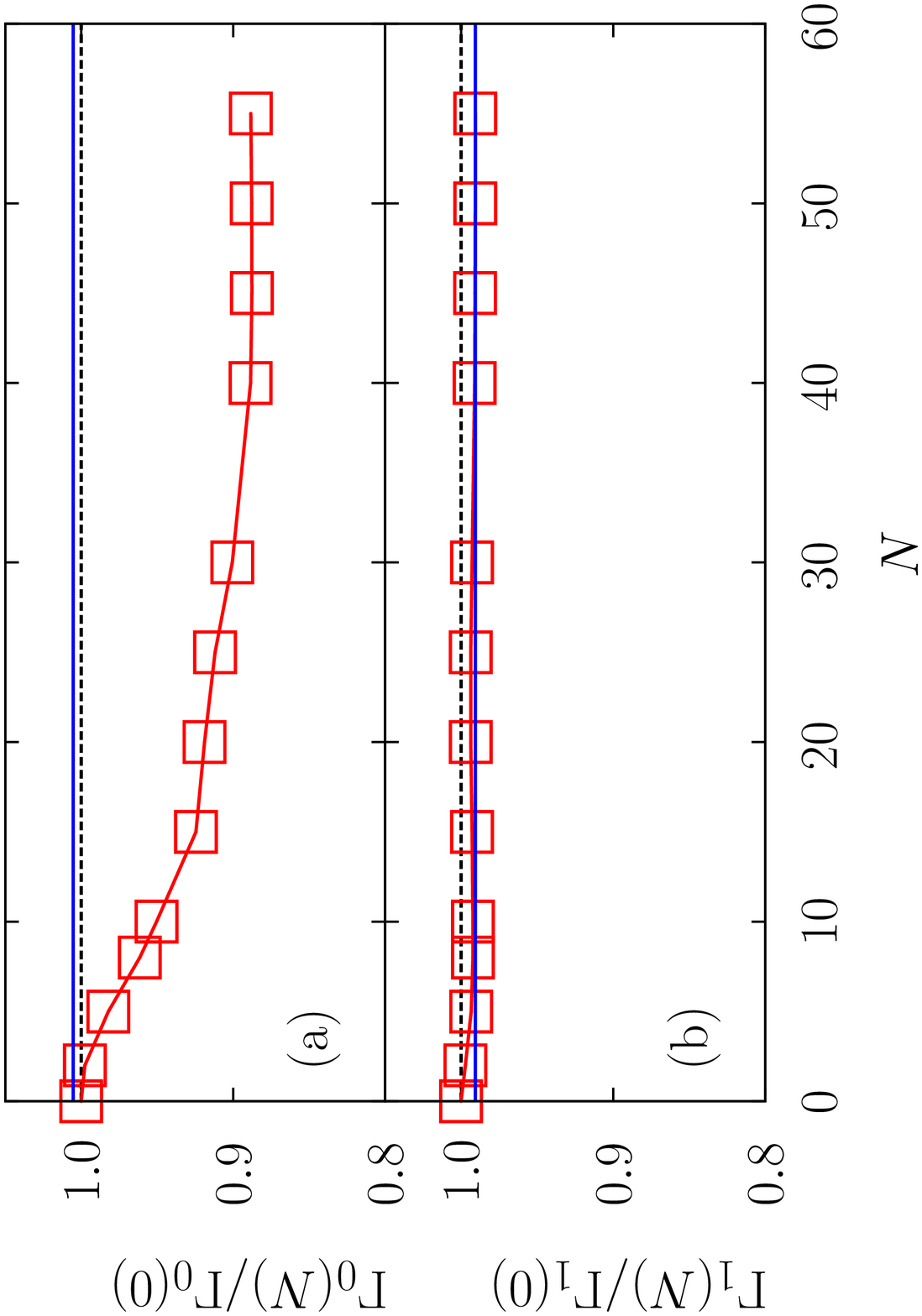}
\caption{(Color online) $N$-dependences of the widths of the states (a) $l = 0$ and (b) $l=1$ in the CdGM mode for the cutoff $l_{\cc} = 3$.  $N$ is the number of modes, counting from the 
lowest-energy mode, in the extended states with positive energies. The type of the vortex is $L_{z}=2$. Solid squares denote the width with $N$ modes. The dashed lines and solid lines denote the width within the single-mode approximation for $l_{\cc} = 3$ and $l_{\cc}  = 30$, respectively, for reference. }
\label{l3wid}
\end{center}
\begin{center}
\includegraphics[width=7cm]{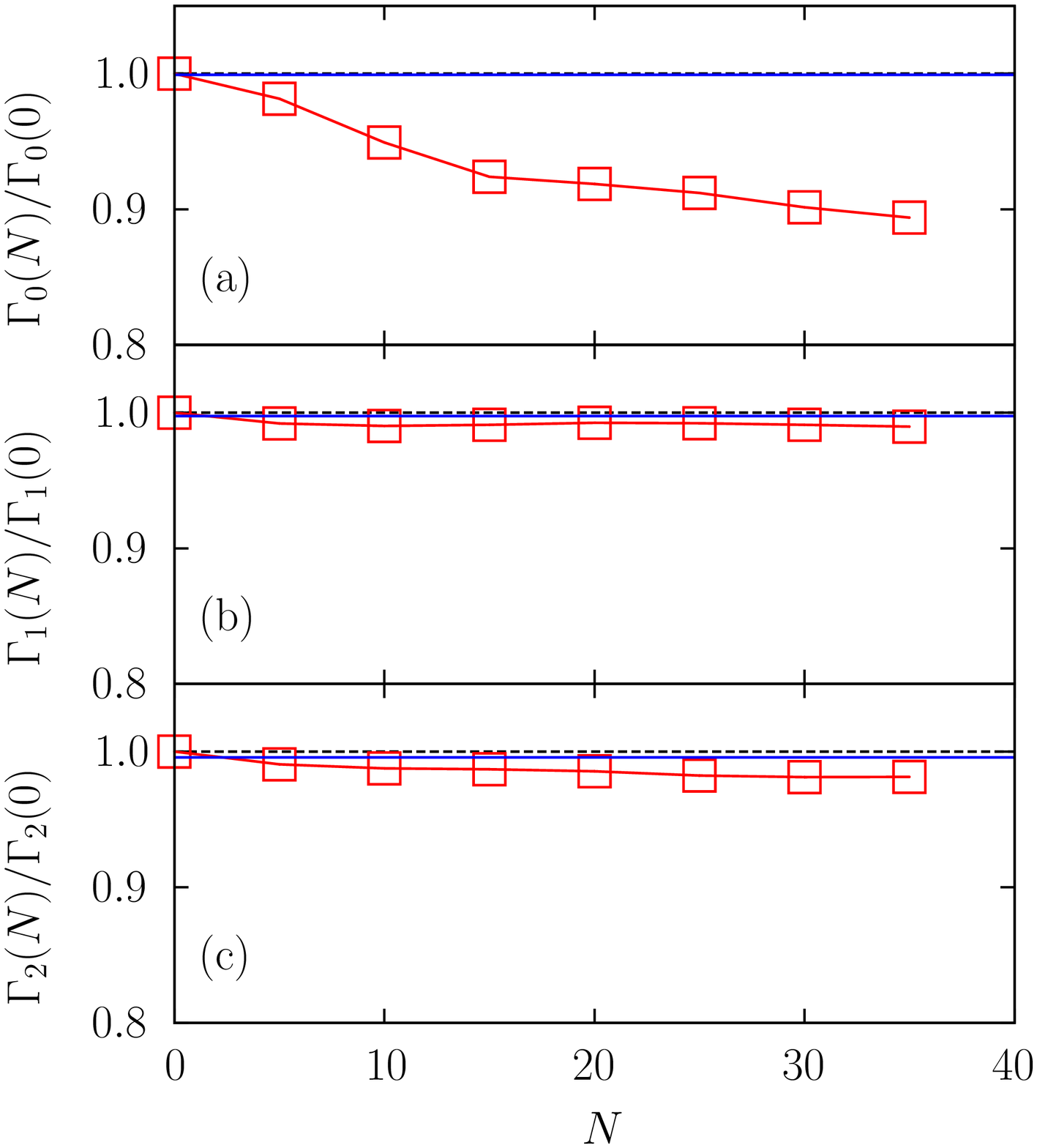}
\caption{(Color online) $N$-dependences of the widths of the states (a) $l = 0$, (b) $l=1$, and (c) $l= 2$ in the CdGM mode for the cutoff $l_{c} = 5$. The symbols are the same as in Fig.~\ref{l3wid}, except that the dashed lines denote the width for $l_{\cc} = 5$.}
\label{l5wid}
\end{center}
\end{figure}
\subsection{Contributions from the modes in the bulk}\label{bulk-mode}
In this section, we evaluate the contributions from the extended states in the bulk with energies above $\Delta_{0}$ 
using Eqs.~\eqref{greenself}--\eqref{defDOS}. 
We consider the chiral p-wave SC with $L_{z}=2$ and use $k_{\F}\xi_{0}=5$ and $\Gamma_{\n} = 10^{-2}\pi\Delta_{0} $. To reduce numerical costs, we set the cutoffs $l_{\cc}$ to $3$ and $5$, where $l_{\cc}$ gives 
the maximum of $|l|$. 
For a given $l$, we take account of $2N + 1$ modes, which consist of $N$ modes with positive energies, $N$ modes with negative energies, and one CdGM mode.
The results for $l_{\cc}=3$ (Fig.~\ref{l3wid}) are almost the same as those for $l_{\cc}=5$ (Fig.~\ref{l5wid}). 
In these figures, open squares represent widths $\Gamma_{i}(N)$ $(i=0,1,2$: the labels of excited states in the vortex core) and 
blue solid lines represent the widths for $N=0$ and $l_{\cc} = 30$. Here, $\Gamma$ used in the above sections corresponds to $\Gamma_{i}(N=0)$.
As $N$ increases, the widths $\Gamma_{i}(N)$ appear to converge.
The contributions to the width of the zero-energy state are about $10\%$ of $\Gamma_{0}(N=0)$ in Figs.~\ref{l3wid}(a) and \ref{l5wid}(a).
Moreover, the contributions to the widths of the first and second excited states with finite excitation energies are less than a few percents of 
$\Gamma_{1}(N=0)$ and $\Gamma_{2}(N=0)$, respectively, in Figs.~\ref{l3wid}(b), \ref{l5wid}(b), and \ref{l5wid}(c).  
There is little effect on excited states; thus, the single-mode approximations are sufficiently valid to obtain the scattering rates. 
\section{Summary and Discussion}
We have developed a scheme based on the Gor'kov Green's function in order to treat the impurity effects on bound states within the self-consistent Born approximation (iKK scheme).
This scheme properly takes account of the coherence factor and thereby it is applicable to various SCs, in contrast to the Kopnin--Kravtsov scheme.
We have confirmed the validity of the iKK scheme by reproducing the result obtained by the quasiclassical theory 
for an s-wave SC and a chiral p-wave SC with $L_{z}=2$ when the system is in the quasiclassical regime. 
For a chiral p-wave SC with $L_{z}=0$, 
on the other hand, the scattering rates calculated in the scheme are small and qualitatively the same but not  quantitatively the same as those obtained by the quasiclassical theory because the spectra have almost no overlaps and are regarded as discrete; $\Gamma \sim 1/\tau < \Delta_{\mathrm{mini}}$.

This scheme has  three advantages over the quasiclassical theory. 
First, the scheme is applicable to the cases 
where $k_{\F}\xi_{0}$ is small and the quasiclassical theory is invalid. 
In this parameter region, the impurity scattering rates calculated by the scheme are quantitatively different from those calculated by the quasiclassical theory.
Considering the difference in the case of small $k_{\F}\xi_{0}$ and the small discrepancy for $L_{z}=0$ mentioned above, we conclude that the quasiclassical theory is valid when  $\Delta_{\mathrm{mini}}\tau \lesssim 1$.

Second, in contrast to the quasiclassical theory, the iKK scheme is effective for calculating the impurity effects on the zero-energy state of a topological SC such as the chiral p-wave SC in this paper.
As a result, for a parallel vortex in a chiral p-wave SC, the width of the zero-energy state is finite but exceptionally small compared with the widths of other excited states when we cannot regard spectra as continuous. (This property was not known in earlier analysis using quasiclassical theory\cite{JPSJ.71.1721}.) 
The width of the first excited state rather than the small width of the Majorana state is thus important in evaluating the minigap.
$M_{0,0} = 0$ leads to this exceptionally small width when the spectra are discrete, as mentioned in Sect.~\ref{kfxi100}.
We infer that this is associated with the character of the Majorana state. 
Although we can discuss the impurity effects on the minigap and adiabaticity of Majorana states in the vortex core, there remains the issue of the origin of the small width of the zero-energy state.
Even though we consider the extended states in the bulk, the width of the zero-energy state remains finite within the self-consistent Born approximation for random averaged impurities.
It is, thus, not the single-mode approximation that causes the finite width of the zero-energy state.

Third, the scheme will be useful for calculating the impurity effects on the vortex core states in undoped Dirac SCs or SCs for which the derivation of the quasiclassical equation is involved owing to the strong spin orbit coupling or Zeeman energy, provided the system has rotational symmetry. 

Lastly, we briefly remark on a possible future work on the two types of vortex in a chiral p-wave SC in terms of topological classification \cite{PhysRevB.78.195125}. A chiral p-wave SC belongs to the BdG class with broken time-reversal symmetry, the class D, which is one of the ten symmetry classes 
given by Altland and Zirnbauer, obtained by using the random matrix classification \cite{PhysRevB.55.1142}. We find, however, that the two types of single vortex in a  chiral p-wave SC exhibit different impurity effects. Our findings strongly suggest another possible classification in terms of symmetry and/or topology in the real space.

\section*{Acknowledgements}
This work was supported by the Program for Leading Graduate Schools, the Ministry of Education, Culture, Sports, Science and Technology, Japan, and by JSPS KAKENHI Grant Numbers
23244070 and 15K05160.

\bibliographystyle{apsrev4-1}
\bibliography{reference}

\end{document}